\documentclass[prl,aps,twocolumn,noshowpacs,noshowkeys,superscriptaddress]{revtex4-1}

\usepackage{amsfonts}
\usepackage{amsmath}
\usepackage{amssymb}
\usepackage{graphicx}
\usepackage{hyperref}

\usepackage{textcomp}%
\newcommand {\bra}[1]{\langle{#1}|}
\newcommand {\ket}[1]{|{#1}\rangle}
\newcommand{\vv}[1]{\mathbf{#1}}

\begin{document}

\def\neel{ Institut N\'eel, CNRS et Universit\'e Grenoble Alpes, 38042 Grenoble, France}
\def\cotton{
Laboratoire Aim\'e Cotton, CNRS, Universit\'e Paris-Sud and ENS Cachan, 91405 Orsay, France}
\author{B.~Pigeau}
\affiliation{\neel}
\author{S.~Rohr}
\affiliation{\neel}
\author{L.~Mercier de L\'epinay}
\affiliation{\neel}
\author{A.~Gloppe}
\affiliation{\neel}
\author{V.~Jacques}
\affiliation{\cotton}
\author{O.~Arcizet}
\affiliation{\neel}
\email{olivier.arcizet@grenoble.cnrs.fr}
\title{Observation of a phononic Mollow triplet in a hybrid spin-nanomechanical system}

\begin{abstract}
{\bf Reminiscent of the bound character of a qubit's dynamics confined on the Bloch sphere, the observation of a Mollow triplet in the resonantly driven qubit fluorescence spectrum represents one of the founding signatures of Quantum Electrodynamics. Here we report on its observation in a hybrid spin-nanomechanical system, where a Nitrogen Vacancy spin qubit  is magnetically coupled to the vibrations of a Silicon Carbide nanowire. A resonant microwave field turns the originally parametric hybrid interaction into a resonant process, where acoustic phonons are now able to induce transitions between the dressed qubit states, leading to synchronized spin-oscillator dynamics. We further explore the vectorial character of the hybrid coupling to the bidimensional deformations of the nanowire. The demonstrated microwave assisted synchronization of the spin-oscillator dynamics opens novel perspectives for the exploration of spin-dependent forces, the key-ingredient for quantum state transfer.}
\end{abstract}
\maketitle

{\it Introduction--} A mechanical oscillator coupled to a two level system (TLS) is a versatile basis to study the interaction between macroscopic and purely quantum objects. This unconventional combination  \cite{Schwab2005,Treutlein2014} is a promising route towards the generation of non-classical states of motion of macroscopic objects. Hybrid coupling signatures have now been demonstrated between a mechanical oscillator and Bose Einstein Condensates \cite{Camerer2011,Joeckel2015}, superconducting qubits \cite{LaHaye2009,O'Connell2010,Pirkkalainen2013},
solid state single spins \cite{Rugar2004,Rabl2009,Arcizet2011,Kolkowitz2012,Bennett2012, Yacoby2012,Ganzhorn2013,Rohr2014,Teissier2014,Ovartchaiyapong2014}, molecules \cite{Tian2014} or quantum dots \cite{Lassagne2009, Steele2009,Sallen2009,Bennett2010,Yeo2013}.\\
\begin{figure}[b!]
\begin{center}
\includegraphics[width=\linewidth]{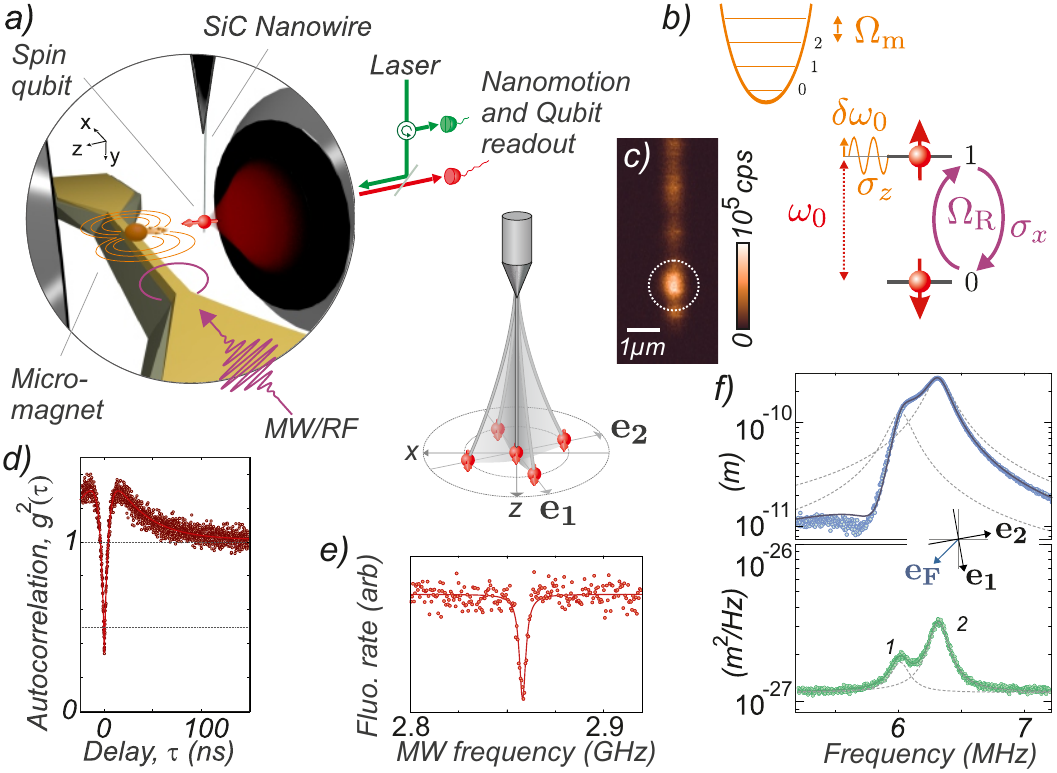}
\caption{
\textbf{ A hybrid spin qubit-nanomechanical system.} {\bf a}. A single NV spin qubit hosted in a diamond nanocrystal is attached at the oscillating extremity of a suspended SiC nanowire.  A strong magnetic field gradient source is micro-positioned in the vicinity of the hybrid system to magnetically couple the spin state to the vibrations of the nanoresonator through the Zeeman effect.  {\bf b.} In the parametric coupling regime, the mechanical motion modulates the qubit energy with an amplitude $\delta\omega_0$.   {\bf c.} Scanning fluorescence image of the hybrid device. {\bf d.} Autocorrelation function of the NV spin qubit fluorescence revealing the presence of a single defect.    {\bf e.} ESR spectroscopy of the suspended spin qubit in a weak magnetic field, highlighting  the spin state dependence of the average emitted fluorescence (FWHM$\sim 4\rm\,MHz$). {\bf e.} Brownian motion of the nanowire measured in reflection (below) revealing the two fundamental eigenmodes, which can be coherently driven through electrostatic or piezo actuations. The response curve (above) permits to determine the local orientation $\vv{e_F}$ of the force vector as well as its magnitude \cite{Gloppe2014}.  }
\label{Fig1}%
\end{center}
\end{figure}
The hybrid interaction coupling phonons and qubits is in profound analogy with quantum electrodynamics (QED) where hallmark experiments revealing the interplay between atoms and photons have permitted exploring the foundations of quantum mechanics. In particular, the apparition of a Mollow triplet in atomic fluorescence spectra \cite{Mollow1969}, characterized by the onset of sidebands appearing on each side of the pump frequency with splitting proportional to the laser field strength, is one of the characteristic signatures of the strongly driven Jaynes-Cumming interaction. Along with the Autler-Townes doublet \cite{Autler1955} or vacuum Rabi oscillations, it expresses the dressing of the atom with the optical photon field \cite{Haroche2006}. Since then, Mollow triplets were observed in atomic vapors \cite{Wu1975, Schabert1975}, single molecules \cite{Wrigge2008,Tamarat1995}, single quantum dots \cite{NickVamivakas2009,Flagg2009} or superconducting qubits \cite{Baur2009} coupled to photon fields in the optical or microwave (MW) domain.
Here we report on the observation of a phononic Mollow triplet, where the phonon field of a nanomechanical oscillator dresses a MW dressed single spin qubit.\\
Our hybrid device consists of a single Nitrogen Vacancy (NV) spin qubit hosted in a diamond nanocrystal attached to the vibrating extremity of a Silicon Carbide (SiC) nanowire \cite{Arcizet2011}. A strong magnetic field gradient couples both components through a spatially dependent Zeeman effect (see Fig.\,1a).
Formally, the generic hybrid spin-oscillator Hamiltonian can be expressed as  $\hbar \omega_0 \sigma_z+\hbar\Omega_{\rm m}(a^\dagger a +\frac{1}{2})+\hbar g_i(a+a^\dagger) \sigma_i$, where $\omega_0/2\pi$ is the qubit energy, $\Omega_{\rm m}/2\pi$  the oscillator frequency, $a^\dagger (a)$ the phonon creation (annihilation) operator, $\sigma_i$ the Pauli matrices of the spin qubit quantized along the $z$ axis and $g_i$ the respective coupling constants. As in quantum electrodynamics, a distinction between the resonant and parametric regimes can be made.
In the first situation, described by the interaction Hamiltonian $\hbar g_{x,y}(a+a^\dagger) \sigma_{x,y}$,  the mechanical oscillator and qubit can coherently exchange single excitations if they have similar frequencies. Several hybrid mechanical systems are exploring this regime, either through a direct interaction \cite{O'Connell2010,Hunger2010,MacQuarrie2013} or mediated by a bus cavity \cite{Pirkkalainen2013,Lecocq2014,Restrepo2014}.
In the case of parametric hybrid coupling, the mechanical motion modulates the qubit energy according to the coupling Hamiltonian $\hbar g_z(a+a^\dagger) \sigma_z$. Reciprocally, the qubit exerts a state dependent force on the oscillator which represents the key ingredient for quantum state transfer between both components. This configuration enables hybrid coupling between components with totally different excitation energies. Moreover by employing a resonant MW tone driving Rabi precession of the qubit at frequencies $\Omega_R/2\pi$ close to the mechanical frequency, it is possible to let both components of the hybrid system evolve on similar time scales \cite{Rohr2014,Saiko2014}. By doing so the parametric interaction with the original qubit is turned into a resonant coupling to the MW dressed qubit. This configuration enables the observation of a phononic Mollow triplet, provided that the phonon field is coherently driven and that the oscillator frequency is larger than the qubit decay rate $\Gamma_{\rm spin}$, which corresponds to the so called resolved sideband regime ($\Omega_{\rm m}>\Gamma_{\rm spin}$) \cite{Teissier2014} of the parametric interaction. In that situation we introduce the dynamical parametric modulation strength $\delta\omega_0$, which denotes the classical amplitude of the mechanically driven parametric modulation (see Fig.\,1b).

The experimental setup is sketched in Fig.\,1a. The nanomechanical oscillator is a  $6\,\mu\rm m$ long SiC nanowire of 300\,nm diameter,   suspended from a sharp metallic tip. Its moving extremity is functionalized with a $\simeq\,50\,\rm nm$ nano-diamond hosting a single NV defect. The hybrid system is investigated with a confocal microscope apparatus (see SI) and a  532\,nm pump laser. It serves for both measuring the vibrations of the nanowire using either the transmitted or reflected light beams and for optical polarization and readout of the qubit  using  spin state dependent fluorescence detection \cite{Jelezko2004}.\\
Measurements of the nanowire Brownian motion permit determining the mechanical properties of the fundamental flexural eigenmodes \cite{Gloppe2014}. These are aligned along two perpendicular directions $\vv{e_{1,2}}$, see Fig.\,1f, tilted by $\approx 10^\circ$ with respect to the optical axis, at frequencies $\Omega_{\rm m}/2\pi$ (m=1 or 2) of 5.99 and 6.29 MHz respectively, with mechanical damping rates $\Gamma_{1(2)}/2\pi=180\, (190)\,\rm kHz$ in air, limited by acoustic emission. The measured effective masses of $M_{\rm eff} \approx 10^{-15}\,\rm kg$ correspond to a spatial spreading of their Brownian motion over $\Delta x_{\rm th}=(k_B T/M_{\rm eff}{\Omega_{\rm m}}^2)^{1/2}\approx 52\,\rm pm$ with zero-point fluctuations of $\Delta x_q\approx 36\,\rm fm$. Using  a resonant force actuation $\vv{\delta F}$, either piezoelectric or electrostatic, it is possible to drive vibrations of the the nanowire around its rest position $\vv{r_0}$. Its vectorial deflection $\vv{\delta r}(t)$ can thus be expressed in frequency space as:
\begin{equation}
\vv{\delta r}[\Omega]=\sum_{{\rm m}=1,2}\chi_{\rm m}[\Omega]\left( \delta  F_{\rm th}^{\rm m}+\vv{\delta F}[\Omega]\cdot \vv{e_m}\right)\vv{e_m}
\label{eq:dr}
\end{equation}
using the mechanical susceptibilities $1/\chi_{\rm m}[\Omega]\equiv M_{\rm eff} ({\Omega_{\rm m}}^2-\Omega^2-i\Omega \Gamma_{\rm m})$ and the independent Langevin forces $\delta F_{\rm th}^{\rm m}$. By adjusting the drive frequency as well as the orientation $\vv{e_F}$ of the force vector with respect to the nanowire eigenmodes it is thus possible  to generate different trajectories in the oscillation plane. This will permit exploring the vectorial character of the magnetic hybrid interaction.\\
The backscattered fluorescence of the NV defect is collected by avalanche photodiodes arranged in a Hanbury Brown and Twiss configuration which allows to confirm the presence of a single NV defect through fluorescence intensity autocorrelation functions (see Fig.\,1d.)\,\cite{MercierdeLepinay2014}. The qubit can be initialized in its ground state through optical pumping, manipulated with quasi-resonant MW fields and readout exploiting its spin state dependent fluorescence rate \cite{Jelezko2004}.  When immersing the suspended NV defect in a strong magnetic field gradient the spin state becomes position dependent, as described by the Zeeman Hamiltonian $-g\mu_B\boldsymbol{\sigma}\cdot \vv{B(\vv{r_0}+\vv{\delta r})}$. At first order the nanomechanical oscillator's vectorial deflection $\vv{\delta r}$ and the spin are dynamically coupled through a magnetic field gradient according to the interaction Hamiltonian:
\begin{equation}
H_{\rm int}=-g\mu_B \left.\boldsymbol{\sigma}\cdot(\vv{\delta r}\cdot\nabla) \vv{B}\right|_{\vv{r_0}}.
\label{eq:Hint}
\end{equation}
Thus, it is generally possible to tune the interaction from resonant ($\sigma_{x,y}$) to parametric ($\sigma_z$) by adjusting the qubit frequency and the topography of the magnetic field gradient. In the following we will only investigate the parametric case, where the nanomotion modulates the qubit energy.

A $18\,\rm\mu m$ diameter NdFeB hard magnetic sphere generates the magnetic field gradient needed to generate a large hybrid coupling strength. It is positioned onto a narrow gold stripline antenna, see Fig.\,2b, used to deliver the resonant microwave field and especially designed to fit between the high numerical aperture objectives without compromising the photon collection efficiency (see SI) and the spatial access to the magnetic gradient source. The latter can be piezo scanned with respect to the suspended NV spin with nanometric precision. The following restrictions dictate the necessary configuration of the experimental setup. Rotating the nanowire while optimizing the collected fluorescence allows to align the intrinsic spin's quantization direction along the optical z axis \cite{Arcizet2011}. The remanent magnetization of the NdFeB particle was oriented in a strong magnetic field ($\approx1.5\,\rm T$) and aligned with the NV natural quantization axis to preserve the spin-selective readout efficiency.
The magnetic bead stray field also allows to polarize the $\rm^{14}N$ nuclear spin by working at the excited state level anticrossing ($\vv{B}\approx\,50\,\rm mT\, \vv{e_z}$) \cite{Jacques2009,Smeltzer2009},  which permits to restrict our system to a pure TLS through MW frequency selection. Finally, the  last fundamental requirement consists in reaching a large parametric coupling strength, orienting  $\boldsymbol{\nabla}B_z$ with the direction of the eigenmode of interest.\\
\begin{figure}[b!]
\begin{center}
\includegraphics[width=0.99\linewidth]{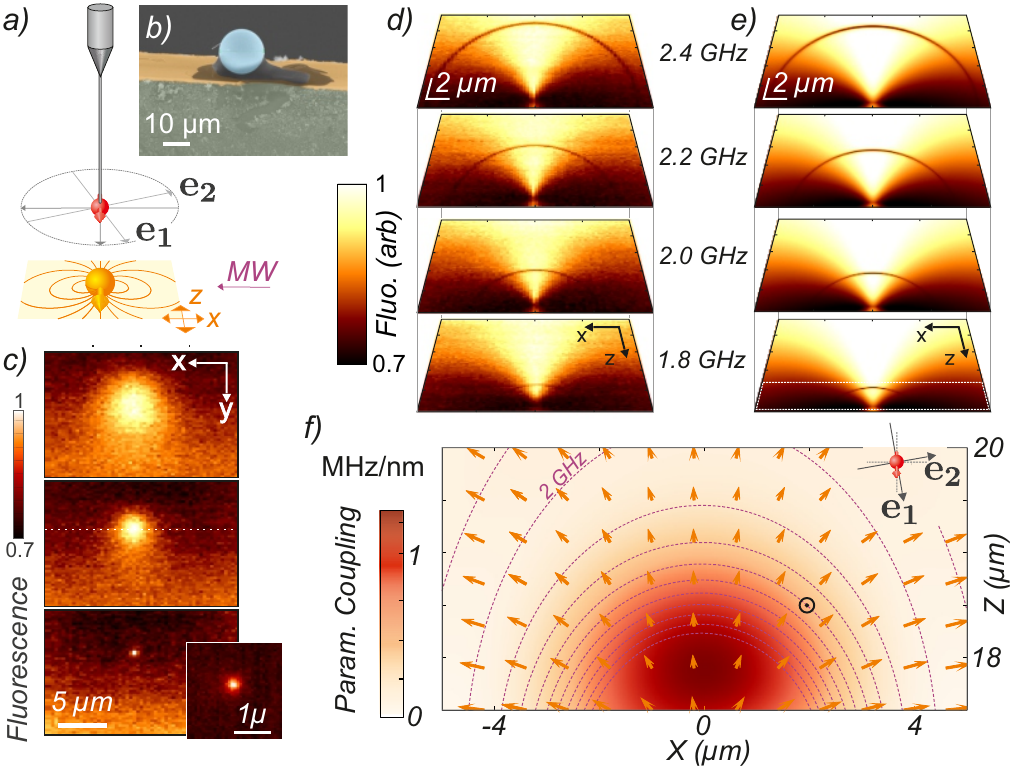}
\caption{
\textbf{ Mapping the parametric vectorial hybrid interaction.} {\bf a.} The coupling strength is determined by measuring the  spatial dependence of the spin qubit energy $\hbar\omega_0(\vv{r})$ in absence of mechanical drive for different positions of the micromagnet. {\bf b.} Colored SEM image showing the $30\,\rm \mu m$ wide suspended MW waveguide supporting the NdFeB microsphere, which is piezo-scanned in the vicinity of the NV. {\bf c.} Fluorescence rates measured at different horizontal positions (22, 20 and 17.5$\,\rm\mu m$) while scanning the micromagnet in the vertical xy plane. The mechanism of fluorescence quenching in transverse magnetic field strongly confines the suitable working points. {\bf d.} Similar measurements in the horizontal xz plane in presence of a single MW tone at varying frequencies. {\bf e.} Corresponding numerical simulations based on a purely dipolar magnetic field (see SI). The resonant slices appear as dark lines which identify the iso-spin energy surfaces $\hbar \omega_0(\vv{r})$, whose spatial derivative permits determining the vectorial parametric coupling strength, $\boldsymbol\lambda\equiv\boldsymbol{\nabla}\omega_0$. Its projection onto the oscillation plane is shown in {\bf f.} as a color map while its orientation is indicated by the orange arrows. The dashed lines reproduce the iso-energy lines (100 MHz spacing). $\odot$ indicates the working point used in the following.}
\label{Fig2}%
\end{center}
\end{figure}
To determine the vectorial parametric coupling strength, the spatial dependence of the spin qubit energy $\omega_0(\vv{r})$ was measured by collecting the NV fluorescence while scanning the position of the micromagnet in presence of a continuous microwave tone. Typical fluorescence maps are shown in Fig.\,2d for varying MW frequencies. The  projections of the qubit iso-energy surfaces on the oscillation plane, see Fig.\,2d, appear as dark
resonant slices \cite{Balasubramanian2008,Degen2009,Maletinsky2012,Rondin2014}. In addition, a global fluorescence quenching which indicates regions with a strong off-axis magnetic field \cite{Tetienne2012}. Reproducing this measurement for varying microwave frequencies permits  determination of $\omega_0(\vv{r})$ (see Fig.\,2f). When moving in those strong magnetic field gradients, the suspended spin qubit undergoes a dynamical parametric energy modulation of $\delta\omega_0(t)=\vv{\delta r}(t)\cdot\left. \boldsymbol{\nabla}\omega_0 \right|_{\vv{r_0}}$, which is determined by evaluating the gradient of the iso-energy map in the ($\vv{e_1},\vv{e_2}$) oscillating plane. The mapping  of the vectorial coupling strength  $\boldsymbol{\lambda}\equiv \left.\boldsymbol{\nabla}\omega_0 \right|_{\vv{r_0}}$  measured in a $10\times2.5\,\rm\mu m^2$ horizontal area in front of the magnetic bead is reproduced in Fig.\,2f. Inherent to the dipolar structure of the microbead magnetic stray field, it strongly varies in magnitude and in orientation $\vv{e_\lambda}$ which permits a fine adjustment of the vectorial coupling strength with respect to the eigenmodes orientations by properly nano-positioning the micromagnet. Furthermore analysis of the fluorescence quenching in this imaging procedure also permits a direct identification of the locations where the $\vv{B}$ field is properly aligned with the NV quantization axis (see Fig.\,2c, 2d),  which is a key requirement to ensure efficient optical spin state readout.
Finally, the triple requirement of avoiding fluorescence quenching, polarizing the Nitrogen nuclear spin and operating with a large parametric coupling strength compete in determining the best location in space where to operate the experiment.\\
\begin{figure}[t!]
\begin{center}
\includegraphics[width=0.99\linewidth]{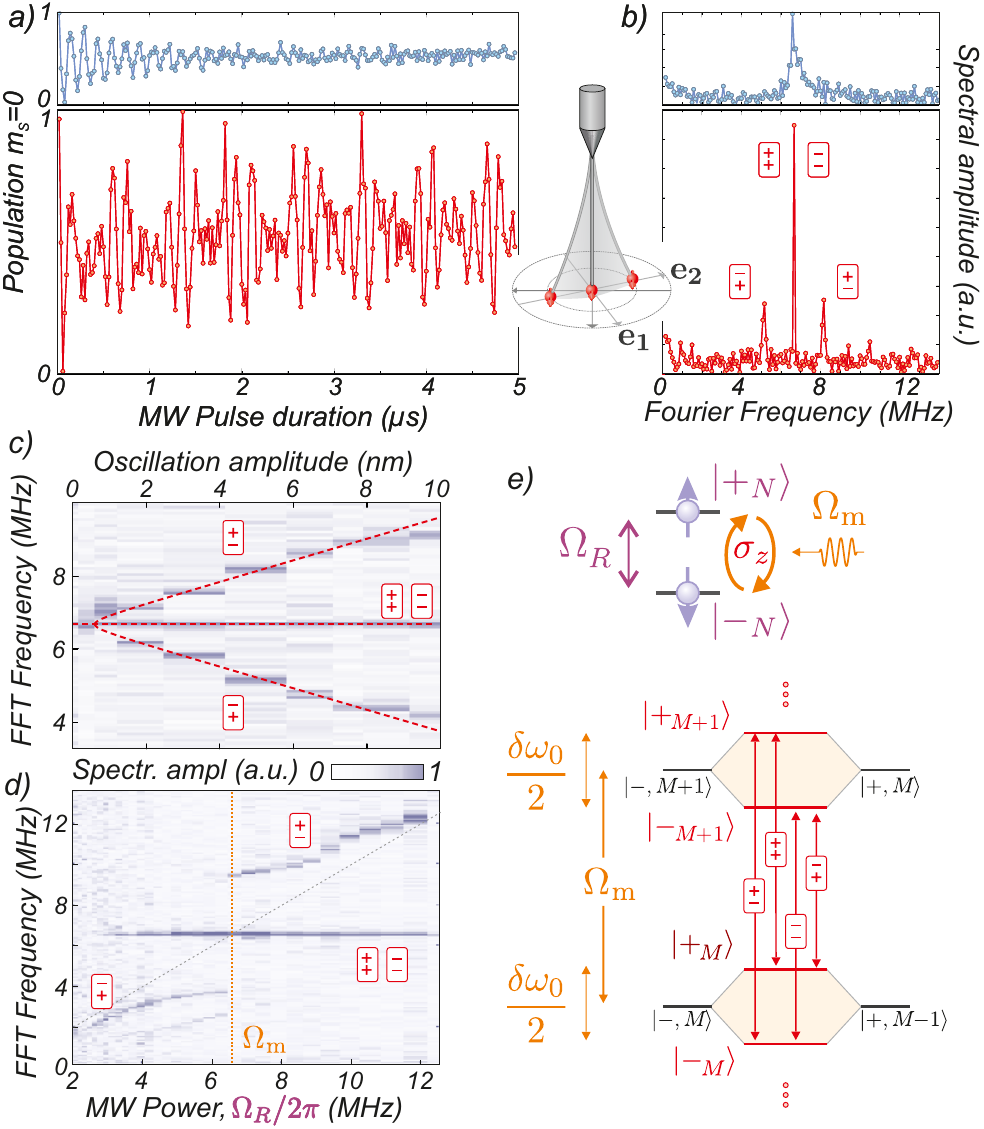}
\caption{
\textbf{ A phononic Mollow triplet.} {\bf a.} Evolution of the spin population when the spin qubit is at rest (top) and oscillating in space (bottom) with a piezo driven 5 nm amplitude along $\vv{e_2}$. {\bf b.}  Magnitude of the corresponding Fourier transforms. A characteristic triplet structure is observed when the spin is oscillating at $\Omega_{\rm d} \approx \Omega_{\rm R}$.  {\bf c.} Triplet separation as a function of the oscillation amplitude. The fit corresponds to a parametric coupling strength of 0.5\,MHz/nm. {\bf d.} Dependence of the triplet structure on the Rabi frequency $\Omega_{\rm R}/2\pi$ detuning.\,\, {\bf e.} The phonon field is dressing the microwave dressed qubit (see text). The measured Rabi oscillations represent a time resolved measurement of the dipole of the dressed qubit, whose spectrum reflects the allowed transitions between  the phonon dressed multiplicities. The Mollow triplet appears when the oscillation amplitude is large enough to create a parametric energy modulation $\delta\omega_0$ that exceeds the spin decay rate. It can be estimated to $\Gamma_{\rm spin}\approx100\rm\,kHz$ from the decay time of Rabi oscillations in the vibrating case.}
\label{Fig3}%
\end{center}
\end{figure}
Having thus fully characterized the static properties of the system, we now investigate the qubit dynamics in presence of coherent mechanical motion, generated by a modulated piezoelectric driving force. We first restrict our analysis to one single mechanical mode ($m=2$) by positioning the gradient source at a location allowing a large parametric coupling strength along the $\vv{e_2}$ orientation and tuning the external drive frequency $\Omega_{\rm d}/2\pi$ to the resonance of the second eigenmode ($6.29\,\rm MHz$). The  qubit is  initialized in its ground state with laser illumination while  the microwave power is adjusted so that $\Omega_R\approx\Omega_{\rm d}$. The subsequent Rabi evolution of the population $\sigma_z(t)$ is shown in Fig.\,3a,  in absence (top) and in presence (bottom) of the coherent mechanical drive. The corresponding Fourier transforms are shown in Fig.\,3b. While first a single decaying oscillation is observed, a longer lasting beating signal can be observed when the qubit is coherently oscillating in space, presenting a characteristic triplet spectrum, whose central component coincides with $\Omega_{\rm d}$, revealing the synchronization of the spin on the oscillator dynamics. Increasing the oscillation amplitude  $\vv{\delta r}[\Omega_{\rm d}]$ from 0 to 9\,nm results in a linear increase of the triplet separation, see Fig.\,3c, up to $\delta\omega_0\approx 2\pi\times 5\,\rm MHz$, corresponding to a 0.5 MHz/nm parametric coupling strength (20\,000 T/m equivalent magnetic gradient), in good agreement with the measured coupling strength $\boldsymbol{\lambda}\cdot\vv{e_2}$ at the target position, see Fig.\,2f. Scanning the MW power permits to illustrate the dependence of the triplet structure in the detuning between $\Omega_R$ and $\Omega_{\rm d}$, and underlines the synchronization of the qubit precession on the driven nanomotion \cite{Rohr2014,Saiko2014}.\\
These observations can be explained by a double dressing of the spin qubit with microwave photon and acoustic phonon fields as follows. The resonant interaction of the microwave pump field with the qubit can be described in the dressed states basis $\ket{\pm_N}$, see Fig.\,3e, parameterized by the number of excitations $N$ shared between the qubit and the MW pump field \cite{Cohen-Tannoudji2004}. Under intense coherent excitation, the dynamics of the spin-microwave  (polariton-like) subsystem can be formally described by a pseudo qubit ($\ket{\pm}$),  see SI, quantized along the microwave polarization axis with a characteristic energy splitting of $\Omega_R/2\pi$ \cite{Cohen-Tannoudji2004}. As a consequence of this rotation of perspective in the Bloch sphere, the respective roles of the $\sigma_{y}, \sigma_{z}$  operators are consequently exchanged. Therefore, the phonon field parametrically coupled to the spin qubit ($\propto(a+a^\dagger)\sigma_z$) is now able to resonantly drive the pseudo qubit, if the resonance condition $\Omega_R\approx \Omega_{\rm d}$ is met.  This second interaction can similarly be described by a second dressing of the pseudo-qubit by the phonon field. This gives rise to a ladder of phonon dressed states, see Fig.\,3e, with eigenstates $\ket{\pm_M}$ parameterized  by the number $M$ of phononic and dressed qubit excitations. The energy splitting within multiplicities can be expressed as (see SI) $\Delta_{\rm Mollow}=\left((\Omega_{\rm d}-\Omega_R)^2+\delta\omega_0[\Omega_{\rm d}]^2/4\right)^{1/2}$, which simplifies to $\delta\omega_0[\Omega_{\rm d}]/2=|\vv{\delta r}[\Omega_{\rm d}]\cdot \boldsymbol{\lambda}|/2$ when the phononic dressing field frequency $\Omega_{\rm d}/2\pi$ is resonant with the dressed qubit energy splitting $\Omega_R/2\pi$. As a consequence the spectrum of Rabi oscillations is peaked at frequencies corresponding to the allowed transitions for the $\sigma_z$ operator (see Fig.\,3e).

Measuring the temporal evolution of the spin qubit population $\langle\sigma_z(t)\rangle$ indeed permits to record the temporal evolution of the dipole of the MW-dressed qubit ($\bra{+}\sigma\ket{-}$) \cite{Cohen-Tannoudji2004}, see SI. This dipole governs the dressed qubit emission (in analogy with the atomic case), whose spectrum  (see Fig.\,3b) reflects the cascade among phononic dressed states. This situation is precisely the one permitting the observation of a Mollow triplet in QED when the atomic fluorescence spectrum under intense illumination was measured \cite{Mollow1969}. An important distinction  is that here the time resolved evolution of the "atomic" dipole (the dressed qubit) is accessible.\\
\begin{figure}[t!]
\begin{center}
\includegraphics[width=0.99\linewidth]{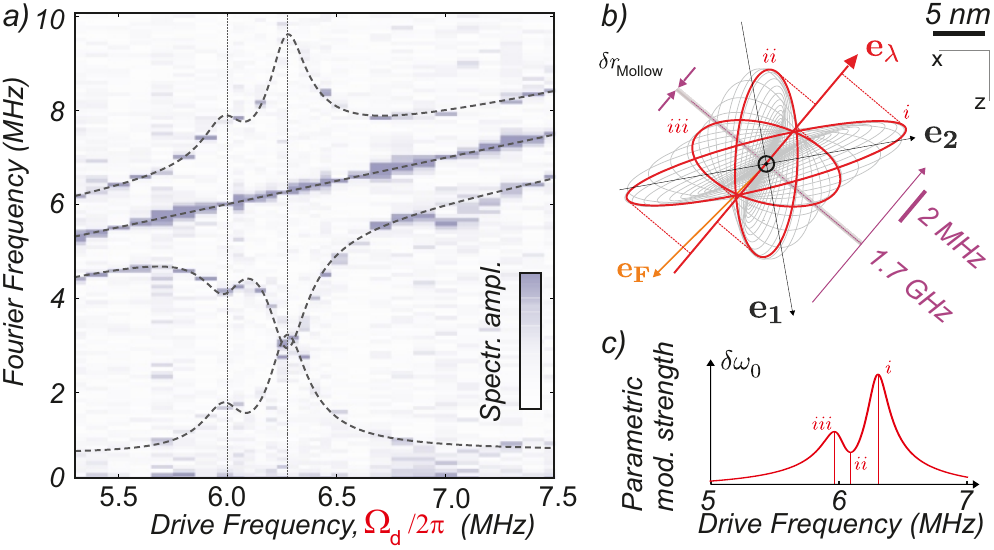}
\caption{
\textbf{ Bimodal Mollow triplet.} {\bf a.} Mollow triplet spectra obtained while sweeping the frequency $\Omega_{\rm d}/2\pi$ of the electrostatic driving force across the two eigenfrequencies, revealing the relative contributions of each eigenmodes ($\vv{e_1},\vv{e_2}$) to the Mollow triplet structure.  The solid lines are $\delta\omega_0[\Omega_{\rm d}]/2,  \Omega_{\rm d}$ and $\Omega_{\rm d}\pm \delta\omega_0[\Omega_{\rm d}]/2$ derived from expression (\ref{eq:domega0}), using the determined eigenmode orientations $\vv{e_{1,2}}$, force orientation $\vv{e_F}$ and the vectorial parametric coupling strength $\boldsymbol{\nabla}\omega_0$ measured at the position marked by $\odot$ in Fig.\,2f. These quantities are reported in {\bf b}  where the spatial trajectories followed by the spin qubit are represented for varying drive frequencies.  Also shown are the qubit iso-energy lines and the characteristic length $\delta r_{\rm Mollow}$ (see text). {\bf c.} The normal projections of the elliptical trajectories on the parametric coupling vector $\vv{e_\lambda}$ determine the Mollow triplet separation. For illustration, the contribution of three specific trajectories are highlighted in panels b and c ({\it i} to {\it iii}). }
\label{Fig3}%
\end{center}
\end{figure}
In order to fully explore the vectorial character of the parametric interaction, we now sweep the drive tone across both mechanical eigenfrequencies. This permits moving the qubit in both directions in the ($\vv{e_1},\vv{e_2}$) oscillating plane. For each drive frequency, the microwave power is adjusted to reach the resonant condition $\Omega_R\approx\Omega_{\rm d}$. The measured Mollow triplet spectra are acquired and shown in Fig.\,4a. The central component of the triplets is locked onto the drive frequency $\Omega_{\rm d}$, while the splitting of the Mollow triplet presents two maxima, corresponding to the response of each eigenmode. In order to properly understand the observed signature, it is necessary to precisely determine the spatial trajectories followed by the moving spin qubit. To do so, an optical measurement similar to the one shown in Fig.\,1c  permits establishing the local orientation  $\vv{e_F}$ and magnitude  $\delta F$ of the electrostatic driving force field \cite{Gloppe2014} and determining the driven trajectories $\vv{\delta r}(t)=Re(\vv{\delta r}[\Omega_{\rm d}] e^{-i \Omega_{\rm d} t})$ using eq.\,(\ref{eq:dr}). The slight spectral overlap between the eigenmodes leads to elliptical trajectories (see Fig.\,4b) which explore the oscillation plane and the magnetic field gradient over nanometric distances. Finally, the Mollow triplet's full splitting can be adjusted with:
\begin{equation}
\delta\omega_0[\Omega]=\left|\sum_{{\rm m}=1,2}\chi_{\rm m}[\Omega]\left( \vv{\delta F}[\Omega]\cdot \vv{e_m}\right)(\vv{e_m}\cdot \boldsymbol{\lambda} )\right|.
\label{eq:domega0}
\end{equation}
Both the deduced  magnitude ($|\boldsymbol{\lambda}|/2\pi=0.5\,\rm MHz/nm$) and the orientation ($\vv{e_\lambda}$ reported in Fig.\,4b)  of the vectorial coupling constant $\boldsymbol{\lambda}$ are in good agreement with the static measurements described above, at the position marked in Fig.\,2f. Geometrically, the magnitude of the parametric coupling strength  corresponds to the length of the projection of the ellipses on the $\vv{e_\lambda}$ axis. Pursuing this geometrical approach, it is possible to introduce a characteristic length, $\delta r_{\rm Mollow}\equiv \Gamma_{\rm spin}/|\boldsymbol{\nabla}\omega_0|\approx 200\,\rm pm$ here (reported in Fig.\,4b), which represents the minimum oscillation amplitude along $\vv{e_\lambda}$ necessary to resolve the phononic Mollow triplet. It is interesting to point out that this quantity is comparable to the spatial spreading of the nanowire Brownian motion of $\approx52\,\rm pm$, responsible for an equivalent incoherent parametric modulation of $\delta\omega_0^{\rm th}=\lambda \Delta x_{\rm th}\approx 2\pi\times 25\,\rm kHz$, which could alter the Mollow triplet structure \cite{Valle2010} in larger magnetic field gradients. Understanding the coherence properties of the dressed qubit and the contribution of Brownian motion will be the subject of future investigation.

{\it Conclusion--}
We have demonstrated the observation of a phononic Mollow triplet in a spin-nanomechanical hybrid system, reproducing with phonons and a spin qubit one of the founding signatures of QED based on photons and atoms. The observed signatures also demonstrate the synchronization of the spin precession onto the mechanical oscillation frequency. This opens the road towards new detection strategies for observing  dynamical spin-dependent forces $\vv{\delta F_{\rm spin}}(t)= \hbar \boldsymbol{\nabla}\omega_0 \sigma_z (t)$. In particular the phononic Mollow triplet structure should also be imprinted on the oscillator dynamics through the spectrum of $\sigma_z[\Omega]$, as mechanical sidebands. Pushing further the analogy with QED \cite{Cohen-Tannoudji2004} it should also provide a direct mechanism for creating single phonon sources through cascaded phonon emission within the dressed state ladder \cite{Wrigge2008}.

\textit{Acknowledgements---} We thank  C. Fabre, G. Nogues, O. Buisson, J.-F. Roch, J.-P. Poizat, P. Vincent, P. Poncharal, A. Auffeves, A. Kuhn,  P. Verlot, E. Dupont-Ferrier, C. Hoarau, D. Lepoittevin and E. Eyraud for theoretical, experimental and technical assistance. This work was supported by the Agence Nationale de la Recherche (RPDoc-2010, FOCUS 2013), Lanef (CryOptics) and the European Research Council (ERC-StG-2012, HQ-NOM). S.R. acknowledges funding from the Nanoscience Foundation.


%

\end{document}